# Computed phase stereo lensless X-ray imaging

J. Duarte[1], R. Cassin[1], J. Huijts[1], B. Iwan[1,2], M. Kovacev[2], M. Fajardo[3], F. Fortuna[4], L. Delbecq[4],

W. Boutu[1] and H. Merdji[1*]

[1]LIDYL, CEA, CNRS and Université Paris-Saclay, CEA Saclay 91191 Gif-sur-Yvette, France.
[2]Leibniz Universität Hannover, Institut für Quantenoptik, Hannover, Germany
[3]Instituto de Plasmas e Fusão Nuclear, IST Lisboa, Portugal
[4]CSNSM, CNRS/IN2P3 and Université Paris-Saclay, 91405 Orsay, France.

[*]correspondence to: hamed.merdji@cea.fr

**The ability to gain insights into the 3D properties of artificial or biological systems is often critical. However, 3D structures are difficult to retrieve at low dose and with extremely fast processing, as most techniques are based on acquiring and computing hundreds of 2D angular projections. This is even more challenging with ultrashort X-rays which allow realizing nanometre scale studies[1–5] and ultrafast time resolved 2D movies[6]. Here we show that computer stereo vision concepts can be transposed to X-rays. We demonstrate nanoscale three-dimensional reconstruction from a single ultrafast acquisition. Two diffraction patterns are recorded simultaneously on a single CCD camera and after phase retrieval two stereo images are reconstructed. A 3D representation of the sample is then computed from quantitative disparity maps with about 130x130x380nm$^3$ voxel resolution in a snapshot of 20 femtoseconds. We extend our demonstration to phase contrast X-ray stereo imaging and reveal hidden 3D features of a sample. Computed phase stereo imaging will find scientific applications at X-ray free electron lasers, synchrotrons and laser-based sources, but also in fast industrial and medical 3D diagnostics.**

In nature, most objects possess complex three-dimensional dynamical structures. The large development of ultrafast coherent X-ray sources allows 2D single-shot imaging on nanometer-femtosecond scale using lensless imaging techniques widely developed on small and large-scale facilities[1-4] but extension to ultrafast 3D imaging is challenging. Nowadays, nanometre scale imaging 3D techniques are mainly based on computed tomography, in which the sample is rotated with respect to the illumination source, allowing for a full set of 2D projections that are recombined to form a 3D image[7,8]. However, such achievement requires hundreds of views. The overall dose is therefore considerably high, effectively damaging biological samples and reducing the achievable temporal and spatial resolutions[9]. To allow imaging extended objects, ptycho-tomography has been proposed[10,11]. While leading to impressive 3D resolutions, this technique is even more demanding, increasing the total acquisition time and the received dose[12]. Imaging before destruction of single particles, as proposed on X-ray FELs, overcomes the radiation dose problem[13]. Nevertheless, this technique requires a huge number of identical samples and generates an extremely large amount of data that need to be sorted, classified and combined to provide a full set of consistent 3D data[14]. There is an intensive work on decreasing the number of orientations, an extreme solution being stereo imaging. Although X-ray stereoscopy was discovered in the end of the 19$^{th}$ century[15], it didn't find wide scientific applications immediately. Recently, however, electron stereopsis microscopy has shown to produce unprecedented 3D perception of nanometre scale details[16]. The main drawback about this approach is that the 3D effect is purely physiological. Indeed, the human brain can get a fast 3D perception of the sample by processing binocular disparities in the cortex region, but without quantitative depth information. Moreover, to allow the cognitive 3D reconstruction, the angle between the two views has to be small, limiting the gain in structural information. Several experiments have taken place at synchrotron beamlines using stereo imaging, but none have achieved a 3D reconstruction stemming from a single shot pair of images[17–19]. In 2008, Schmidt *et al.*[20] proposed a theoretical study of a method dividing an X-ray FEL beam into two sub-beams using a crystal. In 2014, Gallagher-Jones *et al.*[21] probed the 3D structure of an RNAi microsponge by combining *coherent diffractive imaging* (CDI) reconstructions from successive single-shot diffraction patterns from an X-ray FEL, and from X-ray diffraction from synchrotron. However, this method requires several acquisitions to record multiple angles. Techniques to retrieve the 3D structure from a single diffraction pattern have also been proposed, but they work under limited circumstances and heavily rely on sample *a priori* knowledge[22–25]. To date, it has not been possible to obtain a 3D reconstruction from a single X-ray acquisition. Still, stereoscopic coherent imaging has been proposed as a future and promising technique for nanoscale fast time-frame 3D imaging at X-FELs[26]. Here we report a method, based on quantitative disparity map reconstructions, to perform single-shot stereo lensless imaging. This technique extends the *Computer Stereo Vision* from the visible to X-rays using coherent lensless imaging. Instead of constructing a stereo anaglyph with only qualitative 3D perception, our approach retrieves quantitative depth information from two

CDI stereo views. The experimental demonstration is performed using a soft X-ray optical setup based on a high harmonics (HH) beam separated into two coherent twin beams, illuminating a sample with a controllable angle (see Fig. 1). The setup enables recording in a single acquisition two stereo diffraction patterns, reaching nanometre transverse resolution, on a femtosecond timescale, without *a priori* knowledge of the sample. Details on the HH beamline can be found in the Methods section. To generate the two sub-beams, we insert a grazing incidence prism between the off-axis parabola and the sample. Two silicon mirrors are adjusted such that the two beam foci overlap on the sample, with a controllable angle. The setup enables a fine tuning of the time delay with sub-femtosecond jitter between the two pulses and can, alternatively, be used to perform femtosecond time-resolved X-ray-pump/X-ray-probe experiments. The two X-ray beams are diffracted by the sample and the far field patterns are recorded, simultaneously, on a single CCD camera. A typical set of stereo diffraction patterns, acquired at a separation angle of 19°, is shown in Fig. 2. Note that the two patterns present an overlap in the high-spatial frequency regions. However, this does not affect the reconstructions as the useful diffraction information is extracted from a smaller region. The number of useful diffracted photons in each diffraction pattern is roughly equivalent (few $10^7$ photons per shot).

Each diffraction pattern of Fig. 2 is isolated and inverted independently using a HIO "difference map" algorithm[27]. Figs. 3a and 3b show the amplitude reconstructions of the diffraction patterns corresponding to the left and right views, respectively, of the sample of Fig. 3c. They represent the coherent average of 45 independent runs of the phase retrieval algorithm. The spatial resolution of each view is estimated to be 127 nm. Differences between the two views are clear: all the edges of the cross-lid structure are visible in Fig. 3b, whereas some of them are hidden in Fig. 3a.

Qualitative 2D structural and spatial information from two observation angles is achieved in one acquisition. However, it is possible to go further and gather some quantitative depth information from those images. Indeed, from the pair of reconstructed views of the sample, one can compute the disparity map. Disparity refers to the distance between two corresponding points in the two images of a stereo pair. By matching each pixel from one image to the other, one can calculate the distance between them to produce an image where each pixel represents the disparity value for that pixel with respect to the corresponding image. The disparity map can then be converted into depth information by a simple equation, given the geometry of our setup:

$$z(P,\theta) = \frac{d(P)}{tan\theta_1 + tan\theta_2} . \quad (1)$$

In equation (1), $z$ is the relative depth value of the point $P(x,y,z)$ of the object, $d(P)$ is its disparity value and $\theta_1$, $\theta_2$ are the angles between the line perpendicular to the CCD and each stereo beam, respectively. From eq. (1) one can notice that the resolution on the depth axis increases with the angle between the two illuminating beams. However, there is an upper limit for this angle, which is not straightforward to determine as it depends on the samples structure. This limit is defined by a minimal presence of identical features in both views, which are mandatory to calculate the disparity values.

Figs. 4a-b show, respectively, the experimental disparity map and the 3D stereo reconstruction obtained from the image reconstructions shown in Figs. 3a-b. Our geometry leads to a voxel size of 49x49x146 nm$^3$ and an estimated 3D spatial resolution of 127x127x379 nm$^3$. In Fig. 4b the cross is clearly visible, but presents some connections to the membrane. In fact, the sample is a pure-amplitude transmission object, which induces shadow effects as a result of occulted areas in the projection geometry. Moreover, some information is missing due to the lack of common details visible in both views. Surface orientation and edges are, then, under-determined inside shadow regions. Occluding contour artefact is a standard problem in vision science, which can be solved by adding additional 2D views as constraint to surface orientation determination. Multi-view stereo is able to reconstruct detailed 3D information of a scene and is fully exploited in reflective stereo photometry[28]. However, in the context of X-ray vision, and due to optical indexes in this spectral domain, reflectivity of a scene can be quite poor. Instead of multi-view approach, we propose to use phase contrast images that exploit the transparency of a sample to X-rays, available for example at XFELs. Figures 4c and 4d show the results of simulations. Compared to Fig. 4a, which shows an abrupt stop of the disparity values, Fig. 4c shows a continuous sampling of the disparity along the object contour. 2D projections often superimpose objects at different depths on each other, so that subtle differences in the object may not be visible or completely lost. However, by registering two phase contrast images, details can be separated in their x- and y-axis but also along the depth z- axis. A 3D phase stereo reconstruction can be computed and is reported in Figure 4d (see also supplementary movie). The information on the phase shifts unveils the existence of superimposed planes, which makes possible the retrieval of 3D features before hidden behind the membrane. This allows a 3D rendering with fewer artefacts and more details on the structure of the sample.

In conclusion, we have demonstrated a lensless X-ray stereo method that enables quantitative 3D reconstructions with nanometer resolution, in a single acquisition. In our versatile setup, the angle between the two beams can be easily changed, by tilting and moving the plane mirrors, allowing for a geometry adaptation to the sample under study. A larger stereo angle arrangement with a double X-ray detector would allow increasing the numerical aperture of the whole system and, thus, the 3D spatial resolution. An additional advantage of this setup is the possibility to control the temporal overlap between the two beams, at a sub-femtosecond level, enabling 3D X-ray pump – X-ray probe experiments. Moreover, the splitting device can be adapted to shorter wavelength radiation using crystals, thus enabling 3D imaging of fragile biological material with a drastically reduced X-ray dose. Obtaining accurate and realistic disparity maps requires that the two stereo angles have a good overlap between the same details on the object, which could be difficult to obtain with pure-amplitude images. Here, we demonstrated that occulted scenes can be recovered by using phase contrast images as stereo pairs. It considerably lowers the dose delivered to the sample, compared to phase contrast computed tomography which requires hundreds of views. Indeed, extremely fast acquisition times are achieved. For real-time 3D vision at high repetition X-ray sources, such as X-FELs, the image processing can be further increased by using adaptive compressed stereo images, based on redundancies on the two views [29]. We believe that stereo X-ray phase contrast imaging offers tremendous possibilities in biology, materials science or medicine. Furthermore, associated to lensless imaging techniques, it allows following *in vivo* ultrafast 3D motions at a nanometer scale.

**Methods**

**Experimental setup.** The experiment was performed at the SLIC laser centre in CEA Saclay. The high-order harmonic beam setup is described in detail elsewhere[4]. Briefly, we generate the harmonic beam using an amplified Ti:Sa laser system, which delivers 30 mJ, 60 fs pulses, using a loose focusing geometry with a focal length lens of 5.65 m. The XUV beam propagates collinearly with the driving IR laser, which is attenuated using IR antireflective silica plates in grazing incidence. After optimization, we reach $4.10^9$ photons/pulse for harmonic 33 in neon with a spectral bandwidth $\lambda/\Delta\lambda = 150$ and 20 fs pulse duration. A 22.5° off-axis-parabola of $f = 20\ cm$ focal length focuses the harmonic beam to a $5x7\ \mu m^2$ focal spot (FWHM) and selects harmonic 33 ($\lambda = 24\ nm$), thanks to a multilayer coating deposited on its surface. The sample is positioned at the parabola's focus, and the CCD detector (2048×2048 pixels, pixel size $p = 13.5\ \mu m$) is located $z = 26\ mm$ away. Using the sharp edge of the prism, the HH beam is split into two half-beams. Each one is reflected back towards the sample by a pure silicon plate. The prism and silicon plates setup (Fig. 1) is inserted after the parabola in order to increase the angle between the two sub-beams, otherwise limited by the parabola aperture. The focus of each stereo spot is then enlarged (compared to the direct focusing) to 10x7 µm². Note that the whole setup could be placed before the focusing optics, provided that the angle between the two focused beams is large enough. The positions and tilts of the two plates are remotely controlled by vacuum compatible motors, offering the possibility to vary the angle between the two beams. The XUV transmission of the apparatus was estimated to about 75% at a 24 nm wavelength.

**Sample preparation**. The sample is a 6.9x6.1 µm² cross, drilled on a membrane (75 nm of $Si_3N_4$ with a 150 nm Au layer and 4 nm of Cr for adhesion) using a focused ion beam. We first patterned the outer edges of the cross with a low gallium ion current. The soft patterning allows controlling the attachment of the inner cross to the edges. Then, electrostatic forces prevented the lid from falling and "attached" it permanently to the membrane at two opposite contact points.

**Data acquisition and reconstruction.** Although single shot was achieved, we increased the photon flux to use the high dynamic range (HDR) technique. Therefore, two set of diffraction patterns were recorded with integration times of 30 and 140 seconds and recombined. We use a Gaussian filter (σ=2) on the edge of the large-integration time region of interest to stitch it smoothly to the low-range diffraction pattern. Then, we crop any part where there is an overlap between the two diffraction patterns, isolate and reconstruct them using a HIO "difference map" algorithm[27]. We launch 45 independent runs of the algorithm and select the reconstructions that minimize the error criteria. In our case the algorithm converges after roughly a hundred iterations. An image registration algorithm[30] is then used to superpose the best reconstructions and average them.

**Pre-processing of the CDI stereo views.** The computed stereo imaging technique is based on pixel matching. Therefore, a fundamental requirement is that similar illumination conditions apply to both views. In the current configuration, the stereo views, arising from a CDI reconstruction routine, are affected by strong intensity modulations. Those can be due either to artifacts inherent to the CDI algorithm or to experimental issues as for instance the beam's non-uniform intensity profile and partial coherence. Moreover, in a transmission configuration, intensity variations can be ambiguous. Strong intensity attenuations can either arise from density variations on the sample's structures or same-density structures with longer lengths along the imaging axis. To avoid depth calculation errors due to all these phenomena, the followed approach looks for the extraction of depth information along the edged-structures of the sample. The intermediate depth values are then retrieved through 3D interpolation, upon crossing the information of the achieved 3D shape with the amplitude/phase information.

To avoid pixel-matching errors stemming from the causes stated above, the 2D images resultant from the CDI reconstruction are first converted into binary maps. The pre-processing is made as follows. First, the images are resized by a factor 4, with a *bicubic* interpolation in the intermediate pixels. A Gaussian low pass filter is then applied to both images to reduce the effect of the noise ($\sigma$ = 1.9). After filtering, the images are turned binary by defining suited binary thresholds (threshold values in a 0-to-1 scale: 0.40 and 0.25, in left and right images, respectively). Finally, to avoid errors in the binary conversion, the isolated regions constituted by less than 400 aggregated pixels, with no correspondence in the pair image, are removed.

**Computed 3D reconstruction.** Since both views are recorded by the same CCD camera, no image rectification is needed (see Supplementary Section 1.2). The disparity calculations are, then, applied. The disparity maps are calculated employing a simple block matching routine. The images are divided into blocks of 3x3 pixels and, for each picture block, a scan is made over blocks of the same size in the pair picture. The scan is allowed 65 pixels to the left (negative disparity) and to the same amount to the right (positive disparity) of the block central pixel position. A simple sum of absolute differences, added to a less weighted pixel proximity term, is employed as a cost function, to select the best match from the set of candidate blocks (see Supplementary Section 1.3). A second-order sub-pixel interpolation is realized over each disparity value which results from the sorting process. Note that the disparity values are only retrieved over the edges of our 3D structure since it is ambiguous to find matching blocks in the uniform black/white regions. Using this method, two disparity maps are generated, representing the disparity of the left image with respect to the right one (left map, Fig. 4a) and vice versa.

The 3D information is extracted from the disparity maps by employing equation (1), deduced from the camera geometry. For matching the information of the two disparities, a coordinate correction is required. Hence, the $x_1$ and $x_2$ coordinates of the left and right disparity maps, respectively, are converted to the object coordinate $x$. This conversion is obtained from the relations $x(x_1, z, \theta_1) = x_1 - z \tan \theta_1$ (on the left map) and $x(x_2, z, \theta_2) = x_2 - z \tan \theta_2$ (right map).

After retrieving all the coordinates of the points in the 3D space, the stereo consistency of the two disparity maps is evaluated. In this step, the 3D points extracted from each disparity map whose coordinates don't have a match for the second map are discarded.

From the remaining data a point cloud is created and the outlier points are removed. A point is considered an outlier if the average distance to its *k*-nearest neighbours is above a specified threshold *t*. For both experimental and simulation data, the *k* value is defined as 80 points while *t* is 0.1, the latter specifying the number of standard deviations away from the estimated mean distance. The 3D shape of the sample is already visible, with the edged structures completely reconstructed. Next we apply a process in which the information achieved from the 3D reconstruction and the direct stereo views are crossed to achieve the final 3D representation of the sample (see Supplementary Section 1.4).

By fitting a 3D plane in the cross-shape cut of the membrane, a 3D plane surface is computed and a square frame with three points of length is added to the extremities of the point cloud. A 3D scattered interpolation[31] is, then, realized over the resultant point cloud to infer the intermediate values. Crossing the information on the white regions of our stereo views (object transmission function equal to 1), an extra point cloud is computed, composed by stacks of planar points, which we know to correspond to the empty volume of our sample. Excluding from the interpolated 3D mesh the neighbours (0,2 micrometres precision) of the empty region cloud, we reach a final 3D reconstruction of the sample.

**Computed phase stereo imaging process.** Two stereo views are computed with a separation angle of 12° (see Supplementary Section 2.1). To simulate the different phase shifts, distinct absorption values are attributed to the central cross and to the membrane, resulting in stereo views composed by different grey tones. Since the images are not obtained in a parallel camera system, an image rectification step is necessary (see Supplementary Section 1.2). Rectifying a pair of stereo images requires a set of point correspondences between the two views, which is often accomplished by combining feature detection and feature matching algorithms. For an extremely symmetric binary object, however, feature-matching algorithms retrieve ambiguous results, hence a manual choice of matching points is applied. In the simulated stereo images, sixteen edge points are manually selected (see Supplementary Section 2.2). Using all the selected points, the fundamental matrix between both views is computed employing the *Normalized Eight-Point Algorithm*[32]. The images are then re-projected, in order to make all the matching points lay in the same horizontal lines[33].

After the rectification process, the disparity maps are computed, employing the same method used for the experimental data. Besides the two disparity maps obtained from the direct rectified views, two more are calculated. These intend to target specifically the edge areas, allowing for pixel matches in regions which show superposition of different structures in different views. Therefore, a directional gradient along the *x*-axis[34] – direction of disparities - is applied to the rectified stereo views and two new stereo views are generated with a clear highlight on the edges. Two new disparity maps are then extracted from these views, possessing additional information on the structures (see Supplementary Section 2.3). After discarding the inconsistent points between the right and left disparity maps for both cases, the resultant point clouds are merged.

The next step consists in crossing the information of the 3D point cloud and the direct phase images (see Supplementary Section 2.4). Due to the reduced number of views and the existence of superimposed structures, it is necessary to identify isolated sample components and address each component individually. For this step, image segmentation tools and gradient calculations can be used to automatize the process. After identifying the structures, one should fit surfaces in each structure, according with its phase variations and use these surfaces to detect the outlier points. If necessary, some surface points can be added in the active point cloud to help with the 3D interpolation. Note that if the 3D interpolation is made directly for the full point cloud, one can have wrong connections between structures, due to the lack of information in the internal regions.

In the specific case of the simulated sample, since the phase is flat, two planes are fitted to the achieved point cloud. Two new point clouds are then generated, each being constituted by the inlier points of a fitted plane. A 3D scattered interpolation is, then, performed to each cloud and the respective empty volumes removed. Note that all these steps follow the same lines explained for the experimental case. The final reconstruction of the sample is achieved from the assembly of the two 3D structures (Fig. 4d).

**Data availability.** The data that support the plots within this paper and other findings of this study are available from the corresponding author upon reasonable request.

**Acknowledgments**

We acknowledge financial support from the European Union through the VOXEL FET Open H2020 and LASERLAB-EUROPE (grant agreement no. 654148). Support from the French ministry of research through the 2013 ANR grants "NanoImagine", 2014 "IPEX", 2016 "HELLIX"; from the C'NANO research program through the NanoscopiX grant; the LABEX "PALM", through the grants "Plasmon-X" and "HILAC" and, finally, the French ASTRE program through the "NanoLight" grant are also acknowledged.

We acknowledge, as well, the financial support from the Deutsche Forschungsgemeinschaft, grant KO 3798/4-11, and from Lower Saxony through "Quanten- und Nanometrologie" (QUANOMET), project NanoPhotonik.

**Authors contribution**

J.D., R.C., J.H., W.B. and B.I carried out the experiment. The samples were produced by F.F., L.D and W.B.. Simulations were performed by J.D. and R.C.. Data analysis was performed by J.D., R.C., J.H., B.I., M.K. M.F., W. B and H.M.. W.B., M.K. and H.M. proposed the physical concept. All authors discussed the results and contributed to the writing of the manuscript.

**Competing financial interest**

The authors declare no competing financial interest.

**References LETTER**

**References Methods**

**Figures and Figure Legends**

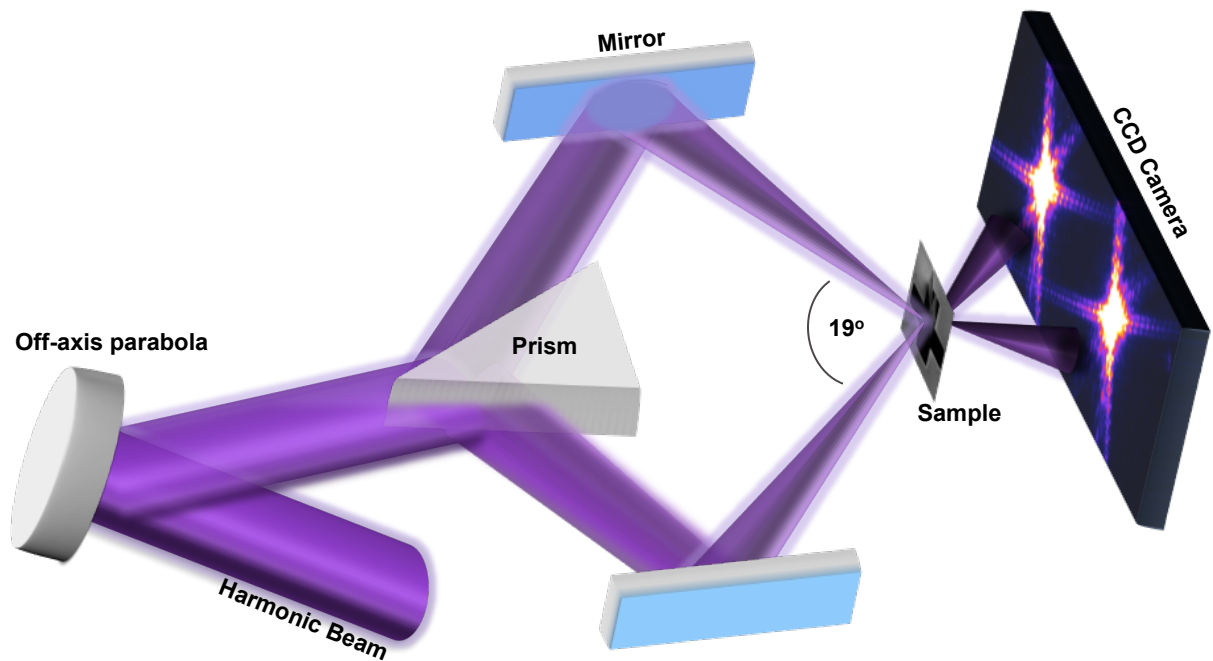

**Fig. 1. Experimental setup for 3D stereo imaging.** A multilayer coated off-axis parabola selects harmonic 33 from the laser (λ=24 nm) and focuses the beam in the sample. A grazing-incidence prism inserted after the focusing optic splits the beam in two. Controllable silicon mirrors are used to reflect each sub-beam onto the sample. A single CCD camera is used to simultaneously record the two diffraction patterns.

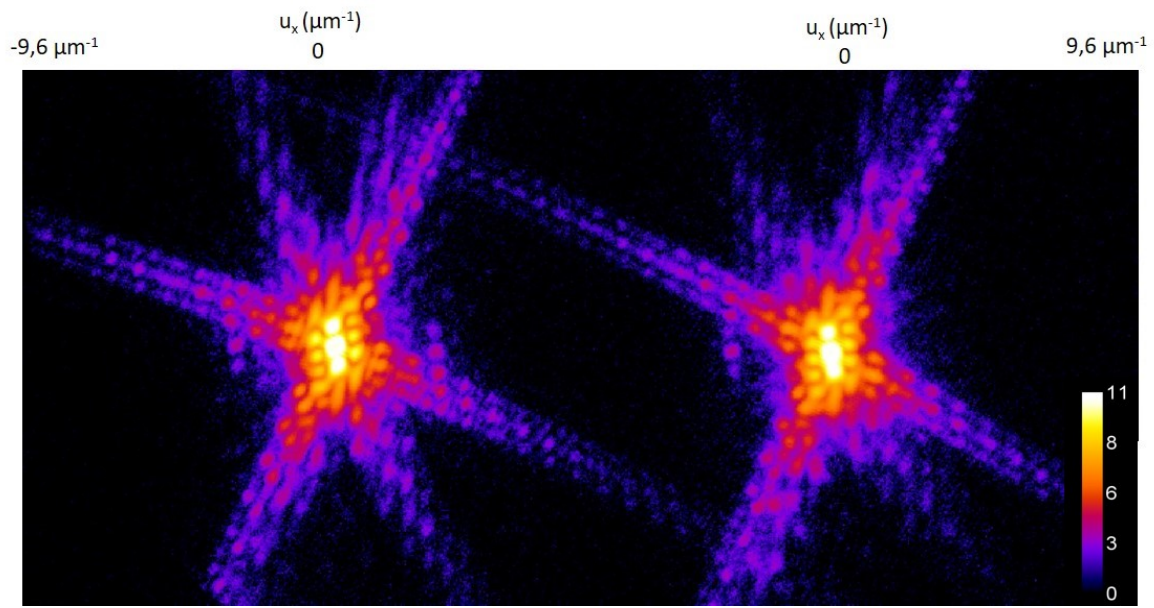

**Fig. 2| Dual diffraction pattern recorded simultaneously on a single X-ray CCD.** The CCD image is shown on a logarithmic scale. The left (right) diffraction pattern corresponds to the beam coming on the sample from the right (left). The diffraction patterns present a slight overlap at high frequencies, which by being in the low-signal part, does not limit the maximal useful diffraction angle. This effect could be circumvented by increasing the stereo angle and using an arrangement with two adjacent CCD cameras or a large-area PN-CCD detector.

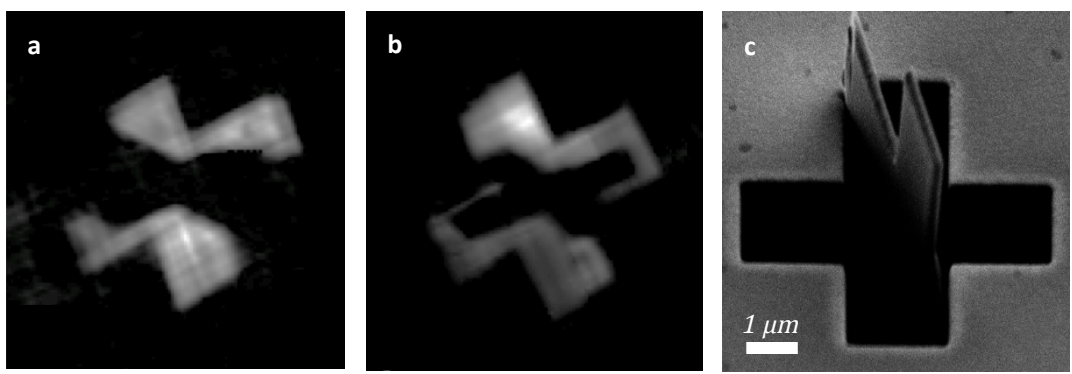

**Fig. 3| 2D amplitude reconstructions of the sample from the two stereo views. a-b,** Reconstructions corresponding to the left and right views of the sample, respectively. They are obtained as the coherent

averages of 45 best reconstructions from independent runs of the CDI algorithm. Each view reaches a spatial resolution of 127 nm which allows to observe details of our nanoscale sample. **c**, SEM (scanning electron microscopy) image of the sample observed at a 60° angle.

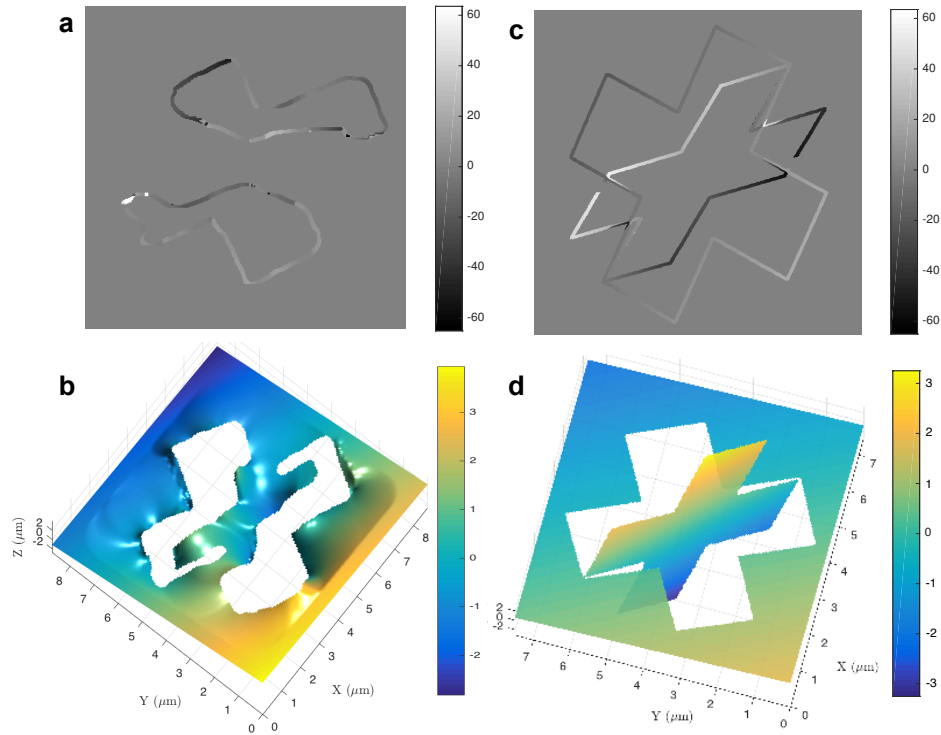

**Fig. 4 | Amplitude and phase computed stereo reconstructions. a-b**, Experimental disparity maps and computed 3D reconstructions of the "cross" sample, respectively**.** The disparity maps are obtained from the correspondence between the left and right views from Fig. 3. The color scale represents the depth value z, for better visualization. **c-d**, Stereo reconstruction of a phase sample using a similar object but adding X-ray transparency. We assume different refraction index materials for the cross and the membrane with a cross-shaped cut. Phase images of each view can be extracted and a disparity map is computed in **c**, from which a 3D reconstruction is achieved, **d**.

15